\documentclass[journal]{IEEEtran}
\usepackage{graphicx}
\usepackage{epsf}
\usepackage{mathtools}
\usepackage{amssymb}
\usepackage{amsmath, amsfonts}
\usepackage{latexsym}
\usepackage{multirow}
\usepackage{multicol}
\usepackage[table]{xcolor}
\usepackage{color}

\usepackage{tabularx}
\usepackage{lipsum}

\usepackage{algorithm}
\usepackage{algorithmic}

\usepackage{mathrsfs}

\usepackage{fixltx2e}

\usepackage[mathscr]{euscript}

\usepackage{commath}
\usepackage{MnSymbol}

\usepackage{subcaption}

\usepackage{mathtools} 
\DeclarePairedDelimiterX\setc[2]{[}{]}{\,#1 \;\delimsize\vert\; #2\,}
\DeclarePairedDelimiterX\parth[2]{(}{)}{\,#1 \;\delimsize\vert\; #2\,}

\PassOptionsToPackage{hyphens}{url}
\usepackage[unicode=true, bookmarks=true, bookmarksnumbered=true, bookmarksopen=true, bookmarksopenlevel=1, breaklinks=false, pdfborder={0 0 0}, pdfborderstyle={}, backref=false, colorlinks=false]{hyperref}

\usepackage{theorem}
{
\theorembodyfont{\rmfamily}

}

\definecolor{orange}{RGB}{255,127,0}
\definecolor{blue}{RGB}{0,0,255}
\definecolor{red}{RGB}{255,0,0}
\definecolor{green}{RGB}{50,160,50}
\definecolor{grey}{RGB}{125,120,125}
\definecolor{purple}{RGB}{125,0,125}

\begin{document}
{
\title{{\fontsize{17}{2}\selectfont On the Feasibility of 4.9 GHz Public Safety Band as Spectrum Option\\\vspace{-0.1 in}for Internet of Vehicles}}

\author
{
Muhammad Faizan Rizwan Khan and Seungmo Kim, \textit{Senior Member}, \textit{IEEE}

\vspace{-0.3 in}

\thanks{M. F. R. Khan and S. Kim are with the Department of Electrical and Computer Engineering, Georgia Southern University in Statesboro, GA, U.S.A. The corresponding author is S. Kim who can be reached at seungmokim@georgiasouthern.edu.}
}

\maketitle
\begin{abstract}
There is an unprecedented impetus on the advancement of internet of vehicles (IoV). The vehicle-to-everything (V2X) communication is well acknowledged as the key technology in constitution of the IoV. Nevertheless, the spectrum for V2X communication is undergoing a massive change in the United States: a majority of the bandwidth has been reallocated to Wi-Fi leaving even less than a half of the bandwidth for V2X. This motivates investigation of other candidate spectrum bands for operation of V2X communication as an urgent effort to guarantee efficient operations of IoV. To this line, this paper studies the feasibility of sharing the 4.9 GHz public safety band between the incumbent systems and V2X users.
\end{abstract}


%%%%%%%%%%%%%%%%%%%%%%%%%%%%%%%%%%%%%%%%%%%%%%%%%%%%%%%%%%%%%%%%%%%%%%%%%%%%%%%%%%%%%%%%%%%%%%%%%%%%%%%%%%%%%%%%%%%%
%%%%%%%%%%%%%%%%%%%%%%%%%%%%%%%%%%%%%%%%%%%%%%%%%%%%%%%%%%%%%%%%%%%%%%%%%%%%%%%%%%%%%%%%%%%%%%%%%%%%%%%%%%%%%%%%%%%%
\section{Introduction}\label{sec_intro}

\subsubsection{Background}
There is no doubt that vehicle-to-everything (V2X) communications have taken the central stage in realization of internet of vehicles (IoV). To date, dedicated short range communications (DSRC) and cellular-V2X (C-V2X) are two key radio access technologies (RATs) that enable V2X communications. Nevertheless, the United States (U.S.) Federal Communications Commission (FCC) voted to significantly reduce the bandwidth for V2X communications as an effort to boost unlicensed operations to support high-throughput broadband applications (viz., Wi-Fi) \cite{sow_5}.

This decision is regarded as serious inefficiency by existing V2X users such as the 50 state departments of transportation (DOTs) in U.S. \cite{sow_6}.

To this line, this paper is dedicated to assessing another spectrum option for the V2X communications.

\subsubsection{Related Work}
This is not the first attempt even for this paper's authors in investigating alternative options for V2X. One of this paper's authors has studied methods of ``\textit{lightening}'' the load of a DSRC network depending on the \textit{crash risk} that each vehicle marks \cite{dave}\cite{access20}\cite{arxiv20}. The method has also been applied to prioritizing urgent operations--viz., military and humanitarian assistance and disaster response (HA/DR)--over civilian operations \cite{milcom19}.

\subsubsection{Contribution}
Distinguished from the related work, this paper (i) clearly identifies a candidate spectrum that has been underutilized and (ii) lays out possible scenarios of operating V2X according to various antenna types.

%%%%%%%%%%%%%%%%%%%%%%%%%%%%%%%%%%%%%%%%%%%%%%%%%%%%%%%%%%%%%%%%%%%%%%%%%%%%%%%%%%%%%%%%%%%%%%%%%%%%%%%%%%%%%%%%%%%%
%%%%%%%%%%%%%%%%%%%%%%%%%%%%%%%%%%%%%%%%%%%%%%%%%%%%%%%%%%%%%%%%%%%%%%%%%%%%%%%%%%%%%%%%%%%%%%%%%%%%%%%%%%%%%%%%%%%%
\section{4.9 GHz Public Safety Band}\label{sec_49ghz}
\subsection{Legislative Background}\label{sec_49ghz_legislative}
In 2002, the U.S. FCC designated 50 MHz of spectrum, i.e., from 4.94 GHz to 4.99 GHz, for the use of public safety purposes. According to the current band plan the band is divided into ten 1 MHz channels and eight 5 MHz channels, while limiting channel aggregation bandwidth to 20 MHz even though, nearly a hundred thousand entities are eligible to be licensed, only around three thousand licenses have been granted by the FCC. The Commission is concerned that the spectrum has not reached its full potential and thus seeks other options to make this spectrum utilized and gain its maximum efficiency.

In the beginning of September 2020, the Commission then voted on the licensing of making the 4.9 GHz widely available for commercial use and allowing to lease some of the rights to third parties either public safety or non-public safety domain, making the spectrum reach its true potential. However, in the summer of 2021 the new administration at FCC halted the process of leasing the spectrum stating that the spectrum shall only be used for the purpose of public safety.

The Code of Federal Regulation (CFR) section 90.523 \cite{cfr90523} defines the eligibility of parties to hold the license for the spectrum, elaborating that all local and state governmental entities are eligible to hold a 4.9 GHz spectrum license, limiting the federal government to not hold the license but allowing them to share it among local and state safety systems. 
The sharing of the systems must be drafted in the form of a legal biding document, explaining in detail the use of the spectrum towards public safety, protection of the lives of citizens, their health or property.

A \textit{band plan} is described as the division of the plan to avoid interference among channels operating in adjacent zones. As per CFR section 90.1213 \cite{cfr901213}, the 4.9 GHz spectrum is permitted to be aggregated among channels comprising of five different bandwidths 5, 10, 15, and 20 MHz. The maximum bandwidth of a channel that could be allocated to a system in 4.9 GHz system is 20 MHz. 

Channels starting from 4940.5 MHz to 4944.5 MHz are 1 MHz allowed bandwidth channels, followed by eight 5-MHz channels up till frequency band 4982.5 MHz, from 4985.5 MHz to 4989.5 MHz are 1-MHz channels as well.

\subsection{Technical Background}\label{sec_49ghz_technical}
\subsubsection{Emission Mask}\label{sec_49ghz_technical_mask}
To maximize the spectral efficiency of the communication over a designated channel the FCC defines spectral emission masks (SEMs). In CFR section 90.210 \cite{cfr90210}, the devices that are being deployed in 4.9 GHz spectrum are required to be in compliance with the FCC-defined masks commonly known as the DSRC-A mask, and the DSRC-C mask. In Fig. \ref{fig_49ghz_mask}, a comparison of existing masks versus the FCC standards is shown along with the attenuation in decibels.

\begin{figure}
\centering
\includegraphics[width = 0.9\linewidth]{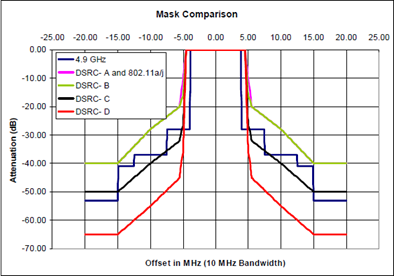}
\caption{Regulation on the 4.9 GHz mask in comparison to related standards \cite{49ghz_mask}}
\label{fig_49ghz_mask}
\end{figure}

\subsubsection{Power Limit}\label{sec_49ghz_technical_power}
According to CFR section 90.1215 \cite{cfr901215}, the transmitting power of stations operational in the frequency band 4940 MHz to 4990 MHz must not exceed the maximum limits. Table \ref{table_power} lists the low-power and high-power maximum conducted power for the aggregated bandwidth of the channels in the safety band spectrum.

\begin{table}[t]
\centering
\caption{Power limits for channel bandwidth 1 MHz - 20 MHz in 4.9 GHz spectrum}
\label{table_power}
\begin{tabular}{| c | c | c |}
\hline
\textbf{Bandwidth (MHz)} & \textbf{Low-power max (W)} & \textbf{High-power max (W)}\\
\hline\hline
1 & 0.005 & 0.1 \\ 
\hline
5 & 0.025 & 0.5 \\
\hline
10 & 0.05 & 1 \\
\hline
15 & 0.075 & 1.5 \\
\hline
20 & 0.1 & 2\\
\hline
\end{tabular}
\end{table}

\subsection{Possible Applications}\label{sec_49ghz_applications}
4.9 GHz spectrum can be utilized using the broadband technology \cite{vtc02}, however, the communication carried over this channel is strictly needed to be done in accordance with the CFR section 90.523 \cite{cfr90523}, which restricts the spectrum to be used for any purpose, but for protection of life, health, or property. 

For instance, video security surveillance can be implemented in a public recreational area and the channel frequency can be set up utilizing the public safety spectrum. In a case of a car crash, or any medical emergency, wireless LAN system can be implemented to clear the scene of incident. Another possible application could be a secure information transmission channel operating at 4.9 GHz that firsthand alerts the hospitals and officials in case of any major threat to the citizens.

%%%%%%%%%%%%%%%%%%%%%%%%%%%%%%%%%%%%%%%%%%%%%%%%%%%%%%%%%%%%%%%%%%%%%%%%%%%%%%%%%%%%%%%%%%%%%%%%%%%%%%%%%%%%%%%%%%%%
%%%%%%%%%%%%%%%%%%%%%%%%%%%%%%%%%%%%%%%%%%%%%%%%%%%%%%%%%%%%%%%%%%%%%%%%%%%%%%%%%%%%%%%%%%%%%%%%%%%%%%%%%%%%%%%%%%%%
\section{Spectrum Sharing in 4.9 GHz Band}\label{sec_sharing}
In consideration of underutilization of the band \cite{carr21}, we propose that the 4.9 GHz band accommodates V2X communications as a \textit{secondary system}. In the following subsections, we identify technical details for (i) the incumbent systems of the band and (ii) spectrum sharing methods of operating V2X communications as a secondary system in the band.

\subsection{Incumbent Systems of 4.9 GHz Band}
Under title 47, Part 2 of CFR \cite{cfr472}, the U.S. spectrum allocation map \cite{map} ranging from frequencies 3.5 GHz to 5.4 GHz lists the authorities that are using the super high frequency (SHF) in the U.S. In particular, near the 4.9 GHz band, the following systems hold an exclusive right for operation.

\subsubsection{U.S. Navy and CEC Training}
Initially, the 4.9 GHz spectrum was allocated to the federal government by the U. S. FCC. However, after 1999 the spectrum higher half of the spectrum was transferred for non-governmental public safety services. Nevertheless, the lower region of 4.9 GHz spectrum is still used by the U.S. Navy as they conduct Cooperative Engagement Capabilities (CEC) training throughout the West, East, Gulf coasts and the state of Hawaii. If licenses are to be leased in the lower half of safety band, they are bound to receive some airborne interference in the areas of operational high power U.S. Navy base stations or in certain situations even 250 miles away from the transmitting towers.

Therefore, any organization that shall seek to get the license for airborne operation in the 4.9 GHz band should file a waiver request, where they could send an application to allow them the use of 4.9 GHz band and modify their operations aligning it with the code defined by the U.S. Navy.

\subsubsection{radio astronomy}
According to the U.S. frequency allocation \cite{map}, the \textit{radio astronomy} is operating in the 4990 MHz to 5000 MHz band on primary basis and operating on 14 specific locations on secondary basis within the range 4950 MHz to 4990 MHz.

Entities leasing the 4.9 GHz are required to operate closely to the ground leaving the operations of radio astronomy without any interference. The same reason goes along with the prohibition of any airborne use of the 4.9 GHz band.

In Fig. \ref{fig_astronomy}, all the radio astronomy sites are shown along with the operational base information in cities around the U.S..

\begin{figure}
\centering
\includegraphics[width = 0.9\linewidth]{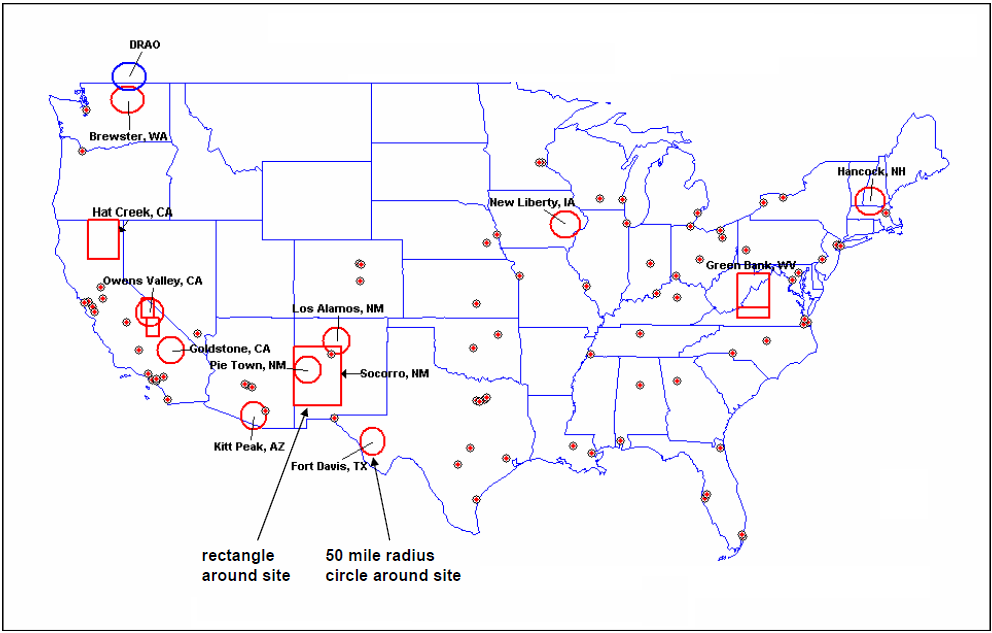}
\caption{Radio astronomy sites in the U.S. \cite{astronomy_map}}
\label{fig_astronomy}
\end{figure}

\subsection{Method of Sharing 4.9 GHz Band with V2X}
The U.S. has been suffering from the number of deaths almost ever increasing ever since the country started releasing the statistics \cite{fatal}. Amongst roads in the nation, the Interstate Highway 16 (I-16) is considered as one of the deadliest routes in the southern part of the U.S. routes \cite{report_5} and recently due to the construction, not only there are traffic congestions all over the routes, but majority of the time there are crashes and during the rainy or hurricane season the whole scenario gets even scary. Considering the aforementioned spectrum sharing setting, establishing a transmitter station (Tx) along the ramp that could help the inbound and outbound traffic with real-time information regarding major accidents or updates along the route of the interchange.

There were two main reasons for choosing the interchange as the Tx location. Firstly, the nearest radio astronomy site is located in Atlanta, Georgia, which is more than 250 miles from the prime location. Secondly, the position of the interchange will serve as a beacon to the traffic for Savannah airport, downtown Savannah, and inbound traffic from I-16 East to Tybee Island and Hilton Head Island.

Furthermore, 5 locations for stationary Rxs are selected around the Tx, which are displayed in Fig. \ref{fig_Tx}. Also notice in the figure that live traffic can also be seen highlighted with a red circle.

\begin{figure}
\centering
\includegraphics[width = 0.9\linewidth]{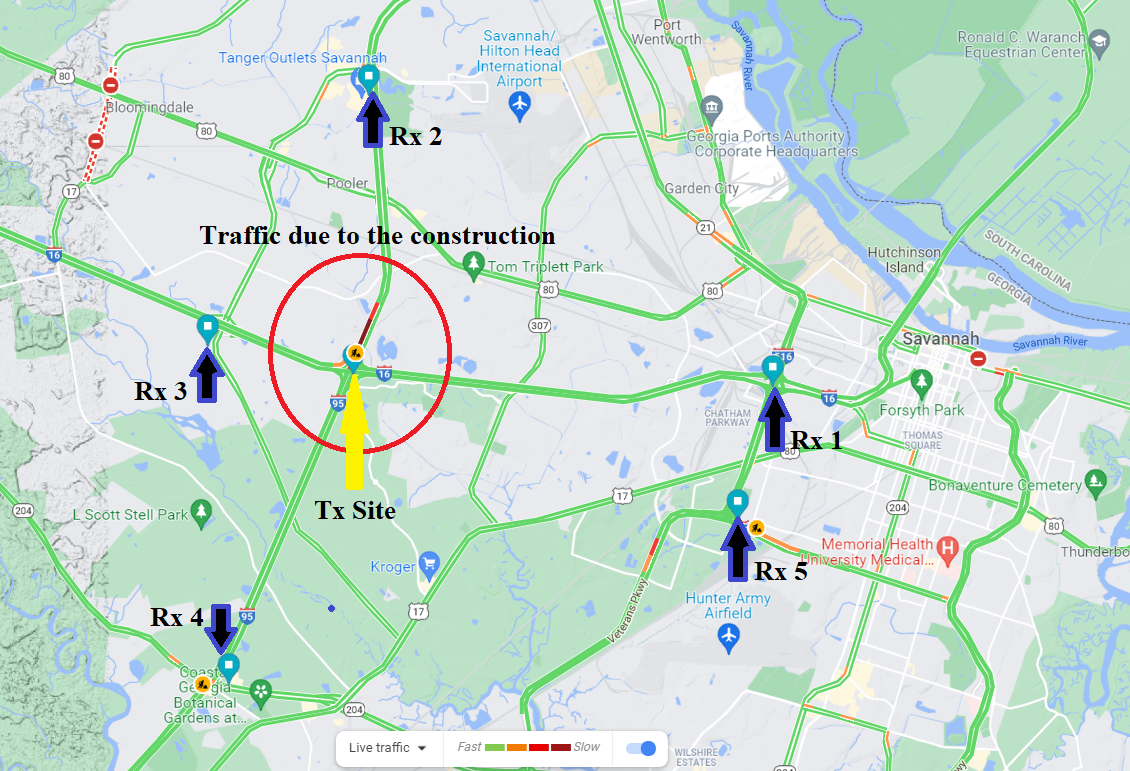}
\caption{Tx site and location of Rxs along with live traffic}
\label{fig_Tx}
\end{figure}

%%%%%%%%%%%%%%%%%%%%%%%%%%%%%%%%%%%%%%%%%%%%%%%%%%%%%%%%%%%%%%%%%%%%%%%%%%%%%%%%%%%%%%%%%%%%%%%%%%%%%%%%%%%%%%%%%%%%
%%%%%%%%%%%%%%%%%%%%%%%%%%%%%%%%%%%%%%%%%%%%%%%%%%%%%%%%%%%%%%%%%%%%%%%%%%%%%%%%%%%%%%%%%%%%%%%%%%%%%%%%%%%%%%%%%%%%
\section{Simulation Results}\label{sec_results}
\subsection{Setup and Parameters}
Key parameters for the simulations are summarized in Table \ref{table_parameters}. By defining the Tx frequency as 4.98 GHz and limiting the power of the Tx by 2 Watts (W) maximum allowed by FCC CFR section 90.1215 \cite{cfr901215}. By dropping the pin over Google Maps, the exact latitude and longitude defining the location are shown.

Communication links will be plotted using the link function along with the ideal coverage of the Tx site. All Rxs should fall ideally inside the boarders of the ideal coverage network. Another useful function used in the simulation is the rain propagation model, which was used to represent the weather condition for the sites.

The central point of this simulation study is a comparison of the antenna radiation types--viz., \textit{dipole} and \textit{directional} antennas. We then proceed to discussing the feasibility of spectrum sharing between the incumbent systems and V2X in the 4.9 GHz band.

\begin{table}[t]
\centering
\caption{Key parameters}
\label{table_parameters}
\begin{tabular}{| c | c |}
\hline
\textbf{Parameter} & \textbf{Value}\\
\hline\hline
Radio access & Long-Term Evoluation (LTE)\\
\hline
Carrier frequency & 4.9 GHz\\
\hline
Bandwidth & 10 MHz\\
\hline
Link & Free space\\
\hline
Tx antenna height & \{60,2\} m \\ 
\hline
Tx power & 1 W \\
\hline
Rx sensitivity & -85 dBm \\
\hline
\end{tabular}
\end{table}

\subsection{Results and Discussion}
Fig. \ref{fig_antenna} compares the coverage of a V2X system near the area of Savannah, Georgia. Figs. \ref{fig_antenna_dipole} and \ref{fig_antenna_directional} represent a Tx adopting a dipole and a directional antenna, respectively.

In Fig. \ref{fig_antenna_dipole}, it can be viewed that all the Rx sites are falling under the umbrella of the Tx site, implying that the simulation received total coverage even when rain possibilities were applied. However, if our receivers (Rxs) were more near the downtown Savannah area it would have been inefficient and the coverage at their end would have outage.

This changes in Fig. \ref{fig_antenna_directional}. With a directional antenna, Rxs 2 and 4 fell out of the coverage. However, the better thing about this antenna is the wide coverage of the directional in the downtown Savannah region. If the project is to be more focused towards a plan that can be deployed in the downtown Savannah region, then it could be more useful.

\begin{figure}
\centering
\begin{subfigure}{\linewidth}
\centering
\includegraphics[width = 0.9\linewidth]{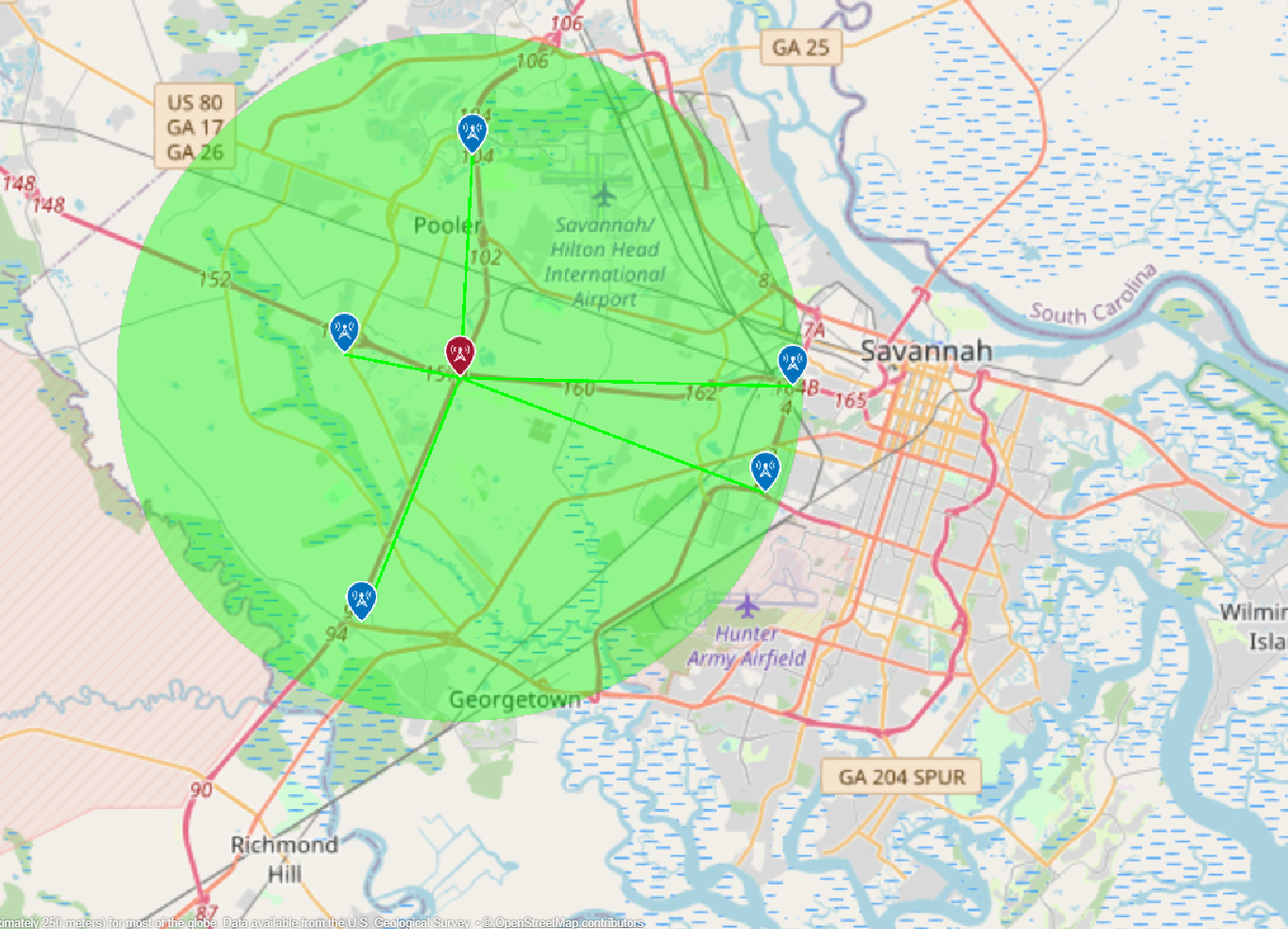}
\caption{Dipole antenna}
\label{fig_antenna_dipole}
\end{subfigure}
\begin{subfigure}{\linewidth}
\centering
\includegraphics[width = 0.9\linewidth]{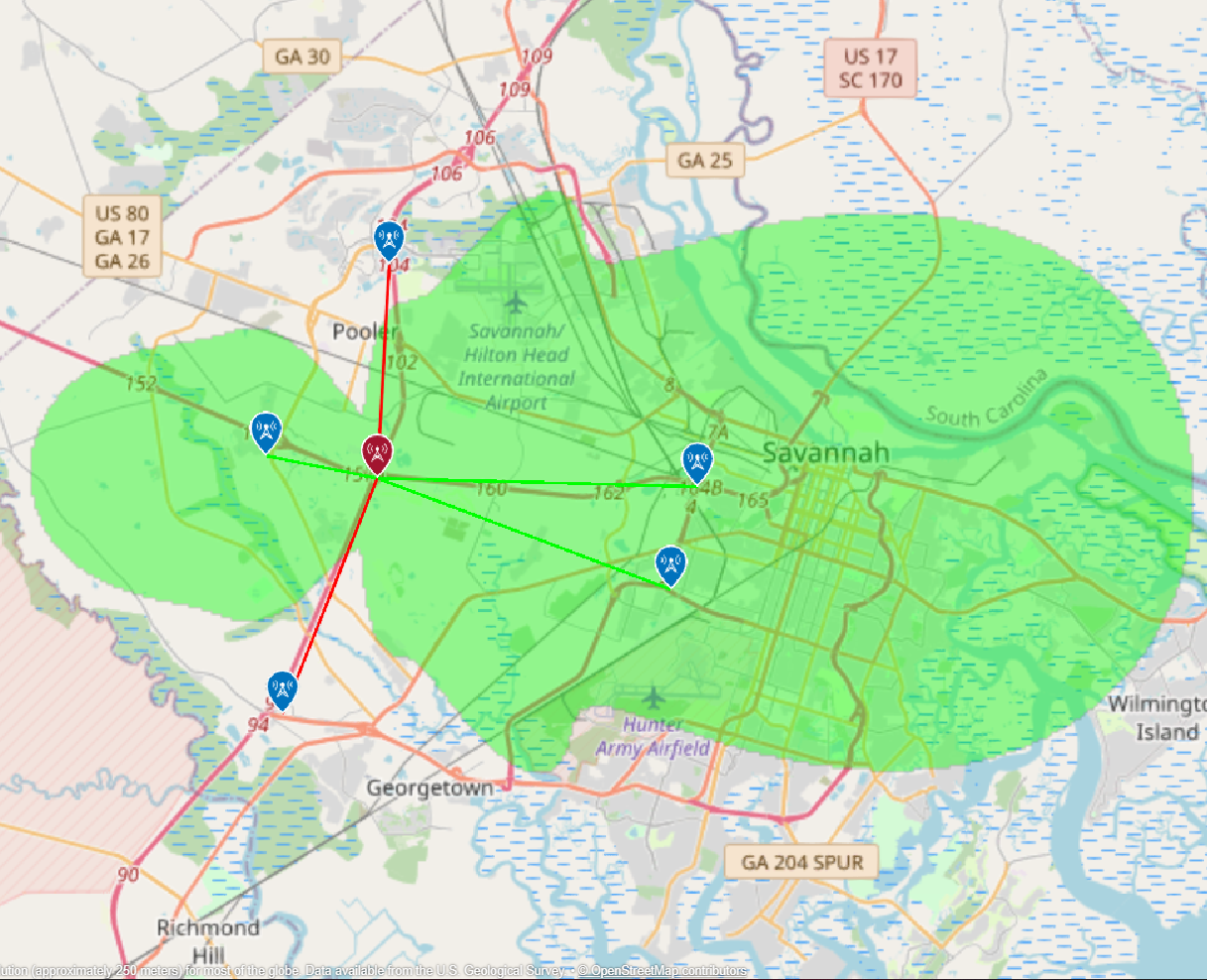}
\caption{Directional antenna}
\label{fig_antenna_directional}
\end{subfigure}
\caption{Comparison of V2X service coverage according to antenna type at Tx (with Tx height of 60 m)}
\label{fig_antenna}
\end{figure}

While a cell tower can easily be as high as 60 m, a lower Tx height makes better sense if a vehicle-to-vehicle, distributed environment is one's mind. Then the concern is a significant reduction in the coverage. As such, we assumed that the Tx antenna is placed \textit{closer} to the Savannah downtown where a higher traffic is observed. Fig. \ref{fig_antenna_2m} illustrates this change. In fact, the Coastal Regional area is with 21,246,000 vehicles traveling through the city of Savannah and nearby areas in 2020 \cite{gdot21}. Further, 65\% of the mentioned traffic was falling in the downtown Savannah area with the directional antenna as displayed in Fig. \ref{fig_antenna_directional_2m}, the simulation is covering from Liberty Parkway Street to Savannah State University a total 5 miles distance. Catering the traffic of the university student's vehicle and tourists, it seems to be an ideal position to implement a Tx away from the incumbent systems existing in the 4.9 GHz spectrum.

\begin{figure}
\centering
\begin{subfigure}{\linewidth}
\centering
\includegraphics[width = 0.9\linewidth]{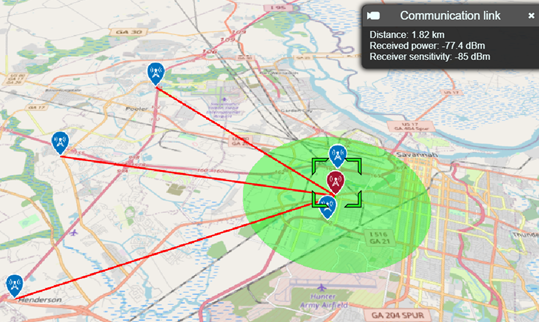}
\caption{Dipole antenna}
\label{fig_antenna_dipole_2m}
\end{subfigure}
\begin{subfigure}{\linewidth}
\centering
\includegraphics[width = 0.9\linewidth]{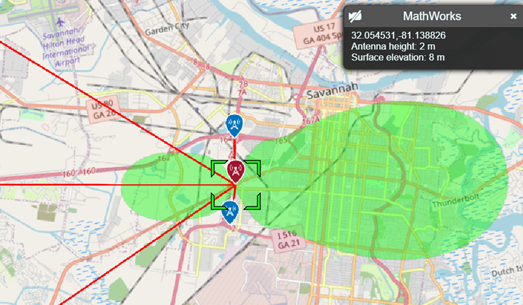}
\caption{Directional antenna}
\label{fig_antenna_directional_2m}
\end{subfigure}
\caption{Comparison of V2X service coverage according to antenna type at Tx (with Tx height of 2 m)}
\label{fig_antenna_2m}
\end{figure}

\subsection{Further Implications}
The model is designed to reduce the traffic congestion near the downtown Savannah region and the Hilton Head International Airport. The simulations provided in this section have already explained the concept behind coverage at the end of Rx and the priority that should be given to the vehicles near downtown area. 

This model can be altered further by cascading several Tx sites throughout the downtown and neighboring suburbs of Savannah, keeping the system operational along with other incumbent system lying in the safety band spectrum. 

In this report, the Tx gave roughly an approximate of 5-mile radius of cell coverage, these approximations were made using a moderate size antenna that could easily be deployed along with traffic lights. Moving along with this approach, around 300,000 residents of Savannah city \cite{report_7} can be benefitted beside the tourist traveling to River Street and nearby beaches along the Southeast coast of U.S.

%%%%%%%%%%%%%%%%%%%%%%%%%%%%%%%%%%%%%%%%%%%%%%%%%%%%%%%%%%%%%%%%%%%%%%%%%%%%%%%%%%%%%%%%%%%%%%%%%%%%%%%%%%%%%%%%%%%%
%%%%%%%%%%%%%%%%%%%%%%%%%%%%%%%%%%%%%%%%%%%%%%%%%%%%%%%%%%%%%%%%%%%%%%%%%%%%%%%%%%%%%%%%%%%%%%%%%%%%%%%%%%%%%%%%%%%%
\section{Conclusions}\label{sec_conclusions}
In this paper, we studied the problem of sharing the 4.9 GHz band between the incumbent systems and V2X users. The findings revealed that the spectrum sharing can be achieved in a space-division manner. Specifically, the usability by V2X users is determined by the incumbent Rx location. In a highly populated area, a directional antenna can improve the possibility of sharing the spectrum between the two systems.

Prospective design insights to IoV can be drawn from this paper's findings. One promising avenue is to apply the proposed spectrum sharing method to an environment where the vehicles are mobile. The dynamicity induced by the mobility will add complexity to the method, which will likely necessitate probabilistic analyses.

%%%%%%%%%%%%%%%%%%%%%%%%%%%%%%%%%%%%%%%%%%%%%%%%%%%%%%%%%%%%%%%%%%%%%%%%%%%%%%%%%%%%%%%%%%%%%%%%%%%%%%%%%%%%%%%%%%%%
%%%%%%%%%%%%%%%%%%%%%%%%%%%%%%%%%%%%%%%%%%%%%%%%%%%%%%%%%%%%%%%%%%%%%%%%%%%%%%%%%%%%%%%%%%%%%%%%%%%%%%%%%%%%%%%%%%%%

\end{document}